\documentclass[prd,nofootinbib,onecolumn,superscriptaddress]{revtex4}

\usepackage[english]{babel}

\usepackage{xcolor,hyperref}
\usepackage{amsmath, amssymb, amsthm, graphicx, epsfig, fancyhdr,epsfig, slashed}

\usepackage{natbib}
\usepackage{float}

\definecolor{lime}{HTML}{A6CE39}

\newcommand{\be}{\begin{equation}}
\newcommand{\ee}{\end{equation}}
\newcommand{\bea}{\begin{eqnarray}}
\newcommand{\eea}{\end{eqnarray}}

\usepackage{feynmp-auto,expdlist}
\usepackage{amsmath, amsfonts, amssymb}
\usepackage{graphicx}
\usepackage{enumerate}
\usepackage{hyperref}
\usepackage{latexsym}
\usepackage{dsfont}
\usepackage{hepnicenames}
\usepackage{enumerate}
\usepackage{soul}
\usepackage[normalem]{ulem}
\usepackage{bbold}

\usepackage{units}

\newcommand{\mio}[1]{}

\def\bpm{\begin{pmatrix}}
\def\epm{\end{pmatrix}}

\usepackage{mathrsfs}

\allowdisplaybreaks
\usepackage{color}
\definecolor{rosso}{cmyk}{0,1,1,0.4}
\definecolor{rossos}{cmyk}{0,1,1,0.55}
\definecolor{rossoc}{cmyk}{0,1,1,0.2}
\definecolor{blu}{cmyk}{1,1,0,0.3}
\definecolor{blus}{cmyk}{1,1,0,0.6}
\definecolor{bluc}{cmyk}{1,1,0,0.1}
\definecolor{verde}{cmyk}{0.92,0,0.59,0.25}
\definecolor{verdec}{cmyk}{0.92,0,0.59,0.15}
\definecolor{verdes}{cmyk}{0.92,0,0.59,0.4}

\newcommand{\beq}{\begin{equation}}
\newcommand{\eeq}{\end{equation}}

\font\tenrsfs=rsfs10 at 12pt
\font\sevenrsfs=rsfs7
\font\fiversfs=rsfs5
\newfam\rsfsfam
\textfont\rsfsfam=\tenrsfs
\scriptfont\rsfsfam=\sevenrsfs
\scriptscriptfont\rsfsfam=\fiversfs

\newsavebox\MBox

\newcommand{\Fc}{\mathcal{F}}

\def\circa#1{\,\raise.3ex\hbox{$#1$\kern-.75em\lower1ex\hbox{$\sim$}}\,}
\makeatletter

\font\ital=cmu10

\def\hhref#1{\href{http://arxiv.org/abs/#1}{arXiv:#1}}
\usepackage{xstring}
\newcommand{\hhrefq}[1]{\IfSubStr{#1}{:}{\href{http://inspirehep.net/search?ln=en&ln=en&p=#1&of=hb&action_search=Search&sf=&so=d&rm=&rg=25&sc=0}{InSpire:#1}}{\hhref{#1}}}

\def\art{\@ifnextchar[{\eart}{\oart}}
\def\eart[#1]#2#3#4#5#6{{\rm #2}, {\em #3 \bf #4} {\rm (#6) #5} ({\em #1})}
\def\article{\@ifnextchar[{\earticle}{\oarticle}}
\def\oarticle#1#2#3#4#5#6{{\rm #1}, {\ital ``#6''}, {\rm #2 #3 (#5) #4}}
\def\earticle[#1]#2#3#4#5#6#7{{\rm #2}, {\ital ``#7''}, {\rm #3 #4 (#6) #5}  [\hhrefq{#1}]}
\def\hepart[#1]#2{{\rm #2, \sl#1}}
\def\heparticle[#1]#2#3{#2, {\ital ``#3''} [\hhrefq{#1}]}
\newcommand{\doi}[1]{\href{http://dx.doi.org/#1}{[link]}}

\def\hhref#1{\href{http://arxiv.org/abs/#1}{arXiv:#1}} 

\begin{document}


\title{Confining complex ghost degrees of freedom}

\author{Marco Frasca}
\email{marcofrasca@mclink.it}
\affiliation{Rome, Italy}

\author{Anish Ghoshal}
\email{anish.ghoshal@fuw.edu.pl}
\affiliation{Institute of Theoretical Physics, Faculty of Physics, University of Warsaw, ul. Pasteura 5, 02-093 Warsaw, Poland}

\author{Alexey S. Koshelev}
\email{ak@inpcs.net}
\affiliation{
Departamento de F\'isica, Centro de Matem\'atica e Aplica\c{c}\~oes (CMA-UBI),\\
Universidade da Beira Interior, 6200 Covilh\~a, Portugal
}

\begin{abstract}

We show a theorem proving that a non-local bosonic field upon a covariant interaction with a confining gauge field undergoes the confinement of its degrees of freedom present in the free theory changing completely the physical mass spectrum following Kugo-Ojima criterion. This is applicable to an infinite number of excitations of the bosonic field including ghosts whereas we pay special attention the the modes with the complex conjugate masses, 
states appearing in the string field theory motivated infinite-derivative models. The same recipe will obviously work for the Lee-Wick models.

\end{abstract}

\maketitle

\textbf{\textit{Introduction:}} Following the Ostrogradski derivation any classical degree of freedom with 4 derivatives can be understood as two degrees of freedom with two derivatives each, inevitably leading to the one possessing negative kinetic energy ~\cite{Ostro}. The latter is known as the ghost. This can be easily generalized to higher but finite number of derivatives. Thinking big one can go for an infinite number of derivatives for example as follows
\be
L=\frac12\phi f(\Box)(\Box-m^2)\phi-V(\phi)
\label{L1}
\ee
Here $f(\Box)$ is often named a form-factor \cite{Efimov:1969fd,Krasnikov:1987yj,Tomboulis:1997gg,Moeller:2002vx,Arefeva:2003mur} and such a construction can be readily tracked for instance in string field theory \cite{Witten:1985cc,Arefeva:2001ps}.
This action is the main staring point of our study and it will be detailed later why and how ghosts may appear here and what is the actual proposal to deal with them.

Another notable example of a model suffering from ghosts is the quadratic curvature gravity presented by Stelle~\cite{Stelle:1976gc} which is renormalizable but not unitary due to the fact that 
the graviton kinetic term has 4 derivatives.
This 4-derivative graviton splits into the massless graviton and a massive ghost with the total propagator
$
1/({M^2_2 p^2 - p^4}) = {M^{-2}_2} [1/p^{2} - 1/({p^2 - M^2_2})]
$
where the minus sign indicates negative kinetic energy.\footnote{During quantization procedure one may look to make positive-energy quantization~\cite{Pais:1950za,Lee:1969fy,hep-th/0503213,astro-ph/0601672,0706.0207,1008.4678,1512.01237,1709.04925,1801.00915},
very similar to the case of fermions (classically the kinetic
energy of fermions is undefined, but a sensible positive-energy quantum theory exists). But then an indefinite norm of states is required \cite{Mannheim:2021oat}.}
This model can be made ghost free \cite{Biswas:2005qr,Koshelev:2013lfm} and renormalizable \cite{Koshelev:2017ebj} by the virtue of a non-local gravity modification such that
\begin{equation}
    S=\int d^4s\sqrt{-g}(\frac{M_P^2}2R+\frac \lambda 2R\Fc(\Box)R+\dots)
\end{equation}
but one still cannot straightforwardly avoid ghost states around all backgrounds \cite{Koshelev:2020fok}\footnote{Certain development attacking this problem can be found in \cite{Modesto:2021soh}}.

In the classical regime classical degrees of freedom having positive kinetic energy which can interact with negative kinetic energy have often run-away solutions, where the individual energies diverge towards large or infinite energies although the total energy remains conserved.
At a first glance ghosts basically may mean
unphysical objects that should be excluded from any sensible theory, that is there should be no ghost-like states in the actual physical spectrum of the theory. But since long it has been known in classical mechanics that several theories containing an interacting ghost have
stable classical solutions with appropriate initial conditions what is known to be ``islands of stability''~\cite{Narnhofer:1978sw,Pagani:1987ue,hep-th/0407231,1302.5257,1607.06589,1703.08929,1811.07733,1811.10019,1902.09557,2003.10860}. 
Whenever the interactions are generic enough and there exists no constant of motion which can forbid the interacting ghost fields to run-away towards dangerous instabilities this situation is manifest. Several numerical studies have revealed that the classical time evolution of ghosts may result in spontaneous lockdown of the ghosts fields, with energies that may vary but remain in a non-trivial restricted range often reaching cosmological meta-stability time ranges
~\cite{Pagani:1987ue}. A perfect example for this is our solar system which is meta-stable, despite no constant of motion forbidding the planets to escape. Considering the classical systems as oscillators with some tiny interactions have been observed to undergo ordered epicycle-like motions, whereas large interactions may lead to chaos as is very known. This means that known physical systems like asteroids in celestial systems and electrons in magnetic fields plus repulsive potentials are described by a ghost degree of freedom, and yet they are meta-stable (see \cite{Gross:2020tph}) leading to existence of ghosts in Nature. Indeed more recently in a study executed in Ref. \cite{Deffayet:2021nnt} the authors presented analytical and numerical proof of existence of ghosts yet the classical motion of the system being completely stable for all initial conditions, notwithstanding that the conserved Hamiltonian is unbounded from below and above. In this paper we going to take a different approach and prove a theorem that the ghosts in presence of size-able interactions gets confined and do not appear as physical states of the mass spectrum, just like quarks and gluons do not appear after QCD confinement.

\smallskip


\textbf{\textit{Appearance of ghosts:}} We turn our attention to action (\ref{L1}). If $f(\Box)$ is present than it must be of the form of an exponent of an entire function in order to have no ghosts in the perturbative vacuum. As such, non-local.
By applying the Weierstrass decomposition which is an infinite product in this case the action can be expanded around any vacuum $\phi_0$ into \cite{Arefeva:2006ido,Koshelev:2007fi,Koshelev:2020fok}
\begin{widetext}
\be
\label{eq:psi}
S=\int d^4x\frac{1}{2}\sum_k\left(\epsilon_k\psi_k(\Box - \omega_k^2)\psi_k
+\bar\epsilon_k\bar\psi_k(\Box - \omega_k^{2*})\bar\psi_k+\mathrm{real~modes}\right).
\ee
\end{widetext}
Here $\omega_k$ are the \textit{complex} eigenvalues of the operator $f(\Box)(\Box-m^2)-V''(\phi_0)$ and $\epsilon_k$ are complex constants while $k$ is just an integer index numbering them.
For a toy $f$ being an exponent of an entire function, like $e^{\Box/M^2}$ with $M$ the non-locality scale, no new states are produced in the perturbative vacuum and the propagator would have no new finite poles. 
However, even though $f$ is finely adjusted for a trivial vacuum, non-trivial vacua of the potential with $V''\neq0$ will immediately lead to an infinite spectrum of states in this case with  complex conjugated masses squared and half of them will look like ghosts. Even if all the vacua have the second derivative of the potential trivial, expansion around the background solution would prompt for an answer of the new modes are.
 The simple but crucial observation here \cite{Koshelev:2007fi,Koshelev:2020fok} is that one cannot avoid an infinite number of states. In case of only complex conjugate masses one is left with a tower of mysterious states which cannot be diagonalized in real values for the masses as noted in the just cited papers. One can make use of the diffusion method to ``localize'' the non-local models in question \cite{Calcagni:2007ef} and as such identify only a finite number of healthy degrees of freedom. Those in turn can be composed out of infinitely many local modes of different nature according to the terminology of the authors. 
The diffusion method implies the absence of ghosts from the physical spectrum, as discussed explicitly in references \cite{Calcagni:2018lyd,Calcagni:2018gke}  where one may find details of the spectrum of the theory at length.


In the present paper we will show that in the quantum approach, captured from quantum field theory (QFT), the classical ghosts with complex conjugate masses coming from higher derivative theories schematically written in (\ref{L1}) with expansion (\ref{eq:psi}) may become confined. This study thus covers the major issue of ghosts in SFT inspired models as well as in analytic infinite derivative gravity theories \footnote{Infinite derivative non-local Higgs has been shown to have very good UV behaviour, like they are Asymptotically Safe and dynamically scale-invariant with a stable vacuum \cite{Ghoshal:2017egr,Ghoshal:2020lfd} and the scale of non-locality affects the vacuum decay \cite{Ghoshal:2022mnj}.} which as has been mentioned above need a treatment for non-trivial backgrounds. Also our analysis should shed more light on that Lee-Wick theories including the generalizations of gravity and the so-called fakeon models originally proposed to tame the fatal ghost degrees of freedom which appear in quantum gravity~\cite{Anselmi:2017ygm} and in Lee-Wick field theories~\cite{Anselmi:2017yux,Anselmi:2017lia,Anselmi:2018kgz} by making them strictly virtual through a different quantization prescription. 


Modes with real masses squared (or simply real modes) in (\ref{eq:psi}) should be restricted to just one mode as other real modes easily become harmful. They are genuine ghosts. This restriction can be achieved by adjusting $f$ but no adjustment of $f$ can eliminate an infinite sum (as $f$ must be an infinite degree polynomial) over eigenmodes completely.

\textbf{\textit{Proof of Confinement:}} Quite recently, confinement for Yang-Mills theory was shown \cite{Chaichian:2018cyv} using Kugo-Ojima criterion \cite{Kugo:1977zq,Kugo:1979gm} (see Appendix A for a short recount of this proof) \footnote{Recently this method was used to study the non-perturbative regimes of higher-derivative theories \cite{Frasca:2022duz,Frasca:2022vvp,Frasca:2021iip}}. In the following, we will show that the field represented in the action (\ref{eq:psi}) are confined when interacting with a confining gauge field. For the time being, we assume the gauge theory to be local. By no means, this implies a loss of generality as a proof of confinement in a non-local infinite-derivative gauge theory was recently presented \cite{Frasca:2021iip} and so, to make an extension is rather straightforward. The entire non-perturbative regimes have been explored in context to high-derivative theories \cite{Frasca:2022duz,Frasca:2022kfy,Calcagni:2022tls,Frasca:2020jbe,Frasca:2020ojd}.


Therefore, we write the above action in Eqn. \ref{eq:psi} adding the interaction term to get the full action (see Appendix B for a derivation)
\begin{widetext}
\be
\label{eq:S}
S_{\texttt total}=\int d^4x\left[-\frac{1}{4}\operatorname{Tr}F^2+
\frac{1}{2}\sum_k\left(\epsilon_k\psi_k(D^2 - \omega_k^2)\psi_k
+\bar\epsilon_k\bar\psi_k(D^2 - \omega_k^{2*})\bar\psi_k\right).
\right]
\ee
\end{widetext}
where we have
\be
F_{\mu\nu}^a=\partial_\mu A_\nu^a-\partial_\nu A_\mu^a-gf^{abc}A_\mu^bA_\nu^c,
\ee
and $f^{abc}$ the structure constants of the group and $g$ the coupling. Similarly,
\be
D_\mu^{ab}=\partial_\mu \delta^{ab}-igA^c_{\mu}(T^{c})^{ab}
\ee
is the covariant derivative with $T^{c}$ the group generators. To action (\ref{eq:S}) we have to add
\be
S_{g}=\int d^4x\left[-\frac{1}{2\xi}(\partial\cdot A)^2-{\bar c}^a\partial^\mu D_{\mu}^{ab}c^b\right]
\ee
being $\xi$ a parameter aimed to fix the gauge and $c^a$ the Faddeev-Popov ghost.

As we know, the following scenario arises in the low-energy limit where confinement is expected is happen \cite{Chaichian:2018cyv,Frasca:2015yva}:
\begin{itemize}
    \item The Faddeev-Popov ghost decouples from the theory.
    \item The gauge field develops a mass gap.
    \item The gauge field becomes confining.
\end{itemize}
Here we consider the approach devised in \cite{Frasca:2015yva} where one solves the Dyson-Schwinger set of equations for the 1P- and 2P-correlation functions \footnote{This approach was utilised to investigate quark confinement in QCD-like theories and hadronic contributions to muon g-2 \cite{Frasca:2022lwp,Frasca:2021zyn,Frasca:2021yuu}}. For our aims, we are interested to the 2P-functions for the Faddeev-Popov ghost and the gluon field. Indeed,
this means that we can write such 2P-functions
in the Landau gauge for the Faddeev-Popov ghost field as
\be
P_2(p)=-\frac{1}{p^2+i\epsilon}
\ee
and for the gauge field, in the same gauge choice, one has
\be
G_2(p)=\frac{\pi^3}{4K^3(i)}
	\sum_{n=0}^\infty\frac{e^{-(n+\frac{1}{2})\pi}}{1+e^{-(2n+1)\pi}}(2n+1)^2\frac{1}{p^2-m_n^2+i\epsilon}.
\ee
We observe that the gauge field 2P-function is in agreement with the K\"allen-Lehman representation with a spectrum of particles given by
\be
m_n=(2n+1)\frac{\pi}{2K(i)}\left(\frac{Ng^2}{2}\right)^\frac{1}{4}\mu,
\ee
where $K(i)$ is the complete elliptical integral of the first kind and $\mu$ is one of the integration constants of the theory considering the background solution to the equations of motion given by the Jacobi Elliptic function as described in details in Ref. \cite{Frasca:2015yva}. At this stage we have neglected the quantum corrections inducing a gap equation for the gauge field \cite{Frasca:2017slg} so, the gauge field propagator is a good approximation to the exact one provided the mass shift induced by quantum fluctuations is small. This is what one sees for SU(N) Yang-Mills theory \cite{Frasca:2017slg}. In the Landau gauge the gauge field 2P-function can be written down in the form
\be
D_{\mu\nu}^{ab}(p)=\delta_{ab}\left(\eta_{\mu\nu}-\frac{p_\mu p_\nu}{p^2}\right)G_2(p),
\ee
where $\eta_{\mu\nu}$ is the Minkowski metric tensor. In the local limit (also Fermi limit as it generates a current-current interaction), this will take the form
\be
\label{eq:Dloc}
D_{\mu\nu}^{ab}(p)=-\kappa\delta_{ab}\left(\eta_{\mu\nu}-\frac{p_\mu p_\nu}{p^2}\right),
\ee
where 
\be 
\kappa = \frac{\pi^3}{4K^3(i)}
	\sum_{n=0}^\infty\frac{e^{-(n+\frac{1}{2})\pi}}{1+e^{-(2n+1)\pi}}(2n+1)^2\frac{1}{m_n^2}
\ee
is a constant.

We can now write out the interaction terms as
\begin{widetext}
\be
S_i=-\int d^4x\sum_k\left[j^a_{k\mu} A^{a\mu}+{\bar j}^a_{k\mu} A^{a\mu}+\frac{1}{2}g^2\psi_k^2A^2+\frac{1}{2}g^2{\bar\psi}_k^2A^2\right].
\ee
\end{widetext}
We have
\be
\label{eq:j}
j_{k\mu}^a=ig\psi_k^b T^a\partial_\mu\psi_k^b,
\ee
and similarly for ${\bar j}_{k\mu}$. 
We just note that the current enters as $j_{k\mu}^a+{\bar j}_{k\mu}^a$. We can write the equations of motion for the gauge field as
\be
D^\mu F^a_{\mu\nu}=-\sum_k\left[
j_{k\mu}^a+{\bar j}_{k\mu}^a+g^2\left(\psi_k^2+{\bar\psi}_k^2\right)A_\nu^a\right].
\ee
We can identify the currents
\be
J_\mu^a=\sum_k\left[j_{k\mu}^a+{\bar j}_{k\mu}^a+g^2\left(\psi_k^2+{\bar\psi}_k^2\right)A_\nu^a\right].
\ee
At a classical level, a functional derivation with respect to the gauge field will yield the equation for the Green function
\be
{\cal L}D_{\mu\nu}^{ab}(x,y)=\delta_{ab}\eta_{\mu\nu}\delta^4(x-y)-g^2
\sum_k\left(\psi_k^2+{\bar\psi}_k^2\right)D_{\mu\nu}^{ab}(x,y)
\ee
being ${\cal L}$ a linear differential operator \cite{Frasca:2015yva}. At the leading order of a functional expansion is
\be
A_\mu^a=\tilde{A}_\mu^a-\int d^4yD_{\mu\nu}^{ab}(x,y)J^{b\nu}(y)
\ee
where
\be
\tilde{A}_\mu^a(x)=\eta_\mu^a\mu\left(2/Ng^2\right)^\frac{1}{4}\operatorname{sn}(p\cdot x+\theta,i),
\ee
where $\eta_\mu^a$ is the polarization vector, $\mu$ and $\theta$ are integration constants and sn a Jacobi elliptical function and such an equation holds provided $p^2=\mu^2\sqrt{Ng^2/2}$ for SU(N). This equation can be solved by iterating. Indeed, for the propagator we get
\begin{widetext}
\be
D_{\mu\nu}^{ab}(x,y)=\tilde{D}_{\mu\nu}^{ab}(x,y)
-g^2\int d^4y'\tilde{D}_{\mu\rho}^{ad}(x,y')
\sum_k\left(\psi_k^2(y')+{\bar\psi}_k^2(y')\right)
\tilde{D}_\nu^{db\rho}(y',y)+\ldots.
\ee
\end{widetext}
Then, we will get
\bea
A_\mu^a&=&\tilde{A}_\mu^a-\int d^4y\tilde{D}_{\mu\nu}^{ab}(x,y)
\sum_k\left[j_{k}^{b\nu}(y)+{\bar j}_{k}^{b\nu}(y)+g^2\left(\psi_k^2(y)+{\bar\psi}_k^2(y)\right)A^{b\nu}(y)\right]+\ldots.
\eea
Due to the mass gap, using eq.(\ref{eq:Dloc}), we can write in the local (Fermi) limit
\be
\label{eq:Dloc2}
D_{\mu\nu}^{ab}(x-y)=-\kappa\delta_{ab}\left(\eta_{\mu\nu}-\frac{\partial_\mu \partial_\nu}{\partial^2}\right)\delta^4(x-y),
\ee
yielding for the interaction term, assuming current conservation,
\be
S_i=\kappa\int d^4x J_\mu^a(x)J^{\mu a}(x),
\ee
provided we take $A_\mu^a(x)=\tilde{A}_\mu^a(x)$. This will generate interaction terms in the form
\be
S'_i=g^4\kappa\int d^4x\sum\left[\psi_k^2(x)+{\bar\psi}_k^2(x)\right]^2\left[\tilde{A}_\mu^a(x)\right]^2
\ee
as the local limit is the relativistic generalization of the London term in superconductivity. Let us note that
\begin{widetext}
\be 
\left[\tilde{A}_\mu^a(x)\right]^2=(N^2-1)\mu^2\sqrt{\frac{2}{Ng^2}}\left[\left(\frac{E(i)}{K(i)}-1\right)+\frac{2\pi^2}{K^2(i)}\sum_{n=1}^\infty\frac{ne^{-n\pi}}{1+e^{-2n\pi}}\cos\left(\frac{n\pi}{K(i)}(p\cdot x+\theta)\right)\right],
\ee
\end{widetext}
where use has been made of the formula $\eta_\mu^a\eta^{a\mu}=N^2-1$ and given the dispersion relation
\be 
p^2=\mu^2\sqrt{\frac{Ng^2}{2}}.
\ee
This series is made by a positive constant plus null-average functions. Indeed, if we choose the rest frame $p=0$ 
the strong coupling limit $Ng^2\rightarrow\infty$ grants rapidly oscillating terms in this series and the situation can only get worse for higher harmonics. This means that a good approximation for the interaction term is
\be 
\label{eq:intt}
S'_i=\frac{\lambda}{4}\int d^4x\sum_k\left(\psi_k^2(x)+{\bar\psi}_k^2(x)\right)^2+\rm{n.a.f.},
\ee
n.a.f. meaning null-average functions and where
\be 
\lambda=\frac{4g^3}{\sqrt{N}}\mu^2\kappa\left(\frac{E(i)}{K(i)}-1\right)
\ee
is the coupling. Now, having a $\phi^4$-like theory, one can apply the analysis performed in \cite{Frasca:2015yva} for the such a theory 
The solution will depend on two integration constants, $\theta$ that is a phase and $\Lambda$ that is an energy-scale. The latter will determine the spectrum that, in this way, will gain complex conjugate poles. This signals that the theory is confining inheriting this property from the gauge field with which is interacting.

\medskip

\textbf{\textit{Discussion and Conclusion:}}

Inspired by string theory, non-local model we investigated (see Eqn. \ref{L1}) with infinite derivatives often displays an infinite number of ghost states. Such ghost states are characterized by having complex conjugate poles with no physical meaning to be attached to them as they carry negative kinetic energy. We proved in our analysis that unlike in free theory if one considers the effect of interaction with a finite coupling which is not necessarily small the situation changes drastically. We proved that such an interaction with a confining gauge theory leads the  mysterious states into the realm of being physical with real masses as the theory acquires a quartic interaction term besides other interaction terms with the gauge field (see Eqn. \ref{eq:intt}). Such a quartic term can be the most important one in the theory if the gauge coupling is taken to be large enough. The reason for this to happen is that the gauge field also displays an infinite number of massive excitations and the interacting scalar fields are driven to behave in the same way by the confining effects of the gauge field which as provided hints for confinement following Kugo-Ojima criterion. It is easy to see that this behaviour is also same as to what happens in quantum chromodynamics (QCD) like theories where the quark confinement occurs. The quark mass can be driven to become unphysical in the confined regime by the gauge degrees of freedom and the quarks and gluonic degrees of freedom and instead replaced by hadronic degrees of freedom \cite{Frasca:2022lwp} again following the Kugo-Ojima criterion. For a similar problem, the interaction term we investigated in Eqn. \ref{eq:intt} to remove the ghosts in the theory was successfully guessed in Ref.~\cite{Galli:2010qx} by one of the authors (A.K.) arriving at similar conclusions as in this paper. The paper considered an exactly solvable model in a gravitational context.
Finally we envisage that our results will shed light on the problem of ghosts in field theory and gravity in general and particularly in higher-derivative and non-local gravity theories. However a detailed analysis in the gravitational framework is beyond the scope of this paper and will be taken up in the future.

\bigskip

While the paper was being finalised we came to know about an ongoing work \cite{Calcagni-new} where they study unitarity and confinement in non-local gravity.

\section*{Acknowledgments}
We thank Gianluca Calcagni and Florian Nortier for valuable comments. AK is supported by FCT Portugal investigator project IF/01607/2015. 

\section*{Appendix A: Confinement in local Yang-Mills theory}
\label{AppendixC}

Confinement in SU(N) Yang-Mills theory was investigated and proven in Ref. \cite{Chaichian:2018cyv}. Utilising the Kugo-Ojima formalism \cite{Kugo:1977zq,Kugo:1979gm}, we have
\begin{widetext}
\begin{eqnarray}
\int d^4xe^{ipx}\langle D_\mu\bar{c}^{a} (x),D_\nu c^{b} (0) \rangle&=&-\delta^{ab}\frac{p_\mu p_\nu}{k^2}\\
&+&\frac{(N^2-1)^2}{2N}g^2\delta^{ab}
\left(\delta_{\mu\nu} - \frac{p_{\mu} p_{\nu}}{p^{2}}\right)
\int\frac{d^4p'}{(2\pi)^4}K_2(p-p')G_2(p').\nonumber
\end{eqnarray}
\end{widetext}
The form of the propagators look like:
\be
K_2(p)=-\frac{1}{p^2+i\epsilon}
\ee
and that of the ghost field and
\be
G_2(p)=\frac{\pi^3}{4K^3(i)}
	\sum_{n=0}^\infty\frac{e^{-(n+\frac{1}{2})\pi}}{1+e^{-(2n+1)\pi}}(2n+1)^2\frac{1}{p^2-m_n^2+i\epsilon}
\ee
and that of the gauge field, 
provided the mass spectrum
\be
m_n=(2n+1)\frac{\pi}{2K(i)}\left(\frac{Ng^2}{2}\right)^\frac{1}{4}\mu,
\ee
where $K(i)$ represents the complete elliptical integral of the first kind and $\mu$ is one of the integration constants of the theory. 
This is a good and natural approximation to the full propagator as long as we omit the mass shift induced due to quantum corrections. Then, the Kugo-Ojima confinement condition looks:
\begin{widetext}
\be
\label{eq:NL}
u(0)=-\frac{(N^2-1)^2}{2N}g^2
\int\frac{d^4p}{(2\pi)^4}\frac{1}{p^2+i\epsilon}\frac{\pi^3}{4K^3(-1)}
	\sum_{n=0}^\infty\frac{e^{-(n+\frac{1}{2})\pi}}{1+e^{-(2n+1)\pi}}(2n+1)^2\frac{1}{p^2-m_n^2+i\epsilon}=-1.
\ee
\end{widetext}
Evaluating the integral one obtains the $\beta$-function in a closed form
\be
\beta_{YM}=-\beta_0\frac{\alpha_s^2}{1-\frac{1}{2}\beta_0\alpha_s},
\ee
with $\beta_0=(N^2-1)^2/8\pi N$. This beta function grants confinement of the theory with the coupling running to infinity at lower momenta with no Landau pole. In the ultraviolet we recover the asymptotic freedom as expected.

\section*{Appendix B: Gauged non-local field theory}

We show how to obtain the action for gauged complex ghosts following the argument given in \cite{Koshelev:2020fok}. Indeed, a gauge invariant non-local field theory, assuming the gauge field being local for reasons of simplicity, takes the form
\be
S_g=\int d^4x\left[-\frac{1}{4}\operatorname{Tr}F^2+\phi^\dagger f(D^2)(D^2-m^2)\phi-V(\phi^\dagger,\phi)\right]
\ee
being $D$ the covariant derivative for the gauge field we are considering. This yields a perfectly gauge invariant, interacting action for a scalar field and a vector field.

We see that, formally, the operator $D^2$ plays the identical role of $\Box$ operator of the free theory and the Weierstrass product decomposition for an entire function applies {\sl mutatis mutandis} in identical manner. Therefore, we expect the decomposition of the operator ${\cal G}(D^2)=f(D^2)(D^2-m^2)$ in the form
\be 
{\cal G}(z)=\prod_i(z-z_i)^{n_i}e^{g(z)}
\ee
being $z_i$ the roots of the function, $n_i$ their multiplicity that we assume to be 1 in the following without losing generality and $g(z)$ is some entire function. The roots are generally complex values. Then, the equation of motion has the form
\be 
{\cal G}(D^2)\phi=\prod_i(D^2-m_i^2)e^{g(D^2)}\phi=0.
\ee
Then, $\phi=\sum_i\phi_i$ where $(D^2-m_i^2)\phi=0$. This extends quite easily to the model we used in the main text taking into account that we have also complex fields.

\end{document}